\documentclass[10pt,twocolumn]{ijcnn}
\usepackage{amsmath}
\usepackage{epsfig}
\usepackage{wrapfig}
\usepackage{bbm}
\usepackage{psfrag}
\usepackage{mymath}
\usepackage{fancyhdr}
\usepackage[mathscr]{eucal}

\graphicspath{{/home/proj_vs/SONN/TAoEC/IEEE/}{/home/proj_vs/SONN/TAoEC/}}

\newcommand{\probdist}{\ensuremath{P}}

\newcommand{\GG}{{\mathcal G}}
\newcommand{\PP}{{\mathcal P}}
\newcommand{\RRR}{{\mathbbm R}}
\newcommand{\gp}{\phi}
\newcommand{\EE}{P_\PP}

\newcommand{\color}[2][1]{} 

\newenvironment{enum}{
\begin{list}{\arabic{enumi}.}{\usecounter{enumi}\leftmargin10pt\topsep1ex\itemsep1ex}
}{
\end{list}
}

\newcommand{\zz}[1]{{\mathcal #1}}

\begin{document}

\title{\Large\textbf{Neutrality: A Necessity for Self-Adaptation}}
\author{\normalsize Marc Toussaint\qquad Christian Igel\\
 \abstractsize Institut f\"ur Neuroinformatik, Ruhr-Universit\"at Bochum\\
\abstractsize 44780 Bochum, Germany\\
\textit{\abstractsize \{\,Marc.Toussaint,Christian.Igel\,\}@neuroinformatik.ruhr-uni-bochum.de}
}
\date{}
\maketitle
\thispagestyle{fancy}

{
\begin{abstract}
  Self-adaptation is used in all main paradigms of evolutionary
  computation to increase efficiency. We claim that the basis of
  self-adaptation is the use of neutrality. In the absence of external
  control neutrality allows a variation of the search distribution
  without the risk of fitness loss.
\end{abstract}
}

\section{Introduction}

It is well known that the genotype-phenotype mapping in natural
evolution is highly redundant (i.e., the mapping is not injective) and
that neutral variations frequently occur. (A variation is called
neutral if it alters the genotype but not the phenotype of an
individual.) The potential positive effects of neutrality have
extensively been discussed
\cite{conrad:90,huynen:96b,kimura:68,schuster:02} and there is growing
interest in investigating how these results from biology carry over to
evolutionary algorithms (EAs)
\cite{barnett:98,ebner:01,igel:01d,newman:98,shackleton:00,shipman:99,weicker:00}.
However, this dissents with the opinion that redundant
encodings are inappropriate for 
\emph{real-world} problems---a bijective genotype-phenotype mapping is
often regarded as a design principle for efficient EAs (e.g., see
\cite{radcliffe:91}).

Unlike neutrality, self-adaptation is an established key concept in
evolutionary computation \cite{baeck:98}. We will point out that
neutrality is a necessity for self-adaptation. And since there is no
doubt about the benefits of self-adaptation for a wide range of
problems, this is a strong argument for the usefulness of neutral
encodings. Our point of view is that, in the absence of external
control, neutrality provides a way to vary the search distribution
without the risk of fitness loss.  We propose to term this way of
adapting the search strategy self-adaptation. This approach
generalizes standard definitions and underlines the central role of
neutrality.

We reconsider two instructive examples from the literature: The first
example is from molecular biology, where it is shown how neutrality
can increase the variability of certain regions in the genome and
conserve the information in other regions.  Using our definition, this
is self-adaptation.  The second one deals with self-adaptation in
evolution strategies, a well-known example of successful
self-adaptation in evolutionary computation.  These algorithms rely on
neutrality to adapt the search strategy and are frequently applied for
solving real-world optimization tasks.

Our arguments are based on recent efforts to provide a unifying view
on the search distribution in evolutionary algorithms and its
\mbox{(self-)}adaptation
\cite{toussaint:01}. 
We introduce this formalism in the following section. Section
\ref{selfada} describes the relation between self-adaptation and
neutrality and in section \ref{examples} we present the two examples
to illustrate our ideas.

\section{Modeling evolutionary exploration}\label{model}

In this section, we outline that the individuals in the population,
the genotype-phenotype mapping and the variation operators including
their parameters can be regarded as a parameterization of the search
distribution. This is done in the framework of global random search
and evolutionary optimization, although all considerations can be
transfered to adaptation in natural evolution, see Sec.~\ref{neu}.

\begin{figure}
\psfrag{Psearch}[r][r]{\small $P_{\mathcal G}=P_{\mathcal P}$}
\psfrag{Pindiv}[r][r]{\small offspring distributions around single individuals}
\psfrag{0.5}[r][r]{\small 0.5}
\psfrag{0.45}[r][r]{\small }
\psfrag{0.4}[r][r]{\small 0.4}
\psfrag{0.35}[r][r]{\small }
\psfrag{0.3}[r][r]{\small 0.3}
\psfrag{0.25}[r][r]{\small }
\psfrag{0.2}[r][r]{\small 0.2}
\psfrag{0.15}[r][r]{\small }
\psfrag{0.1}[r][r]{\small 0.1}
\psfrag{0.05}[r][r]{\small }
\psfrag{-10}[][]{\small -10}
\psfrag{10}[][]{\small 10}
\psfrag{5}[r][r]{\small 5}
\psfrag{-5}[r][r]{\small -5}
\psfrag{0}[r][r]{\small 0}
\centerline{\includegraphics[width=.95\columnwidth]{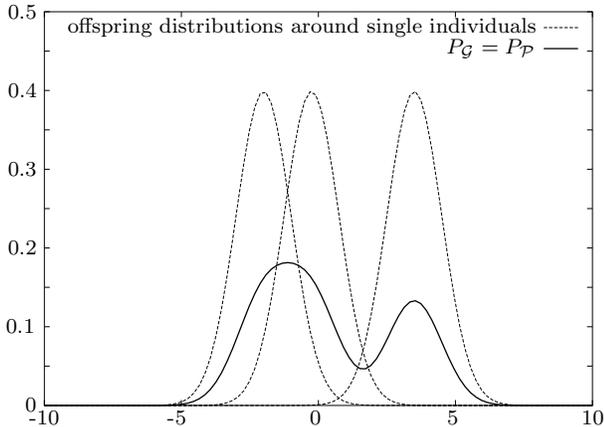}}
\caption{\label{huete:fig}
  An example for parameterizing the exploration distribution in
  evolution strategies: Consider the search space $\PP=\RRR$, $\mu =
  3$ parent individuals, and no recombination. The parents represent
  the object parameters $-2$, $-0.25$, and $3.5$. The offspring is
  generated in the following manner: First, one randomly chosen parent
  is reproduced.  Second, the reproduced individual is mutated by
  adding a realization of a normally distributed random number with
  variance one and expectation zero.  Hence, the resulting search
  distribution (the joint generating distribution
  \protect\cite{rudolph:94}) is a multimodal mixture of Gaussians. }
\end{figure}

Evolutionary algorithms can be considered as a certain class of global
random search algorithms.  Let the search problem under consideration
be described by a quality function $\Phi: \PP \rightarrow \RRR$ to be
optimized. The set $\PP$ denotes the search space.  According to
\cite{zhigljavsky:91}, the general scheme of global random search is
given by:\footnote{This scheme does not account for all kinds of
  evolutionary computation. For example, if the evolutionary
  programming scheme is used, where each parent generates exactly one
  offspring \cite{fogel:95}, or recombination is employed, then the EA
  is better described as operating on a search distribution
  $\probdist^{(t)}_{\PP^\lambda}$ on ${\PP^\lambda}$. However, in
  these cases $\probdist^{(t)}_{\PP}$ can still be derived from
  $\probdist^{(t)}_{\PP^\lambda}$.}
\begin{enum}
\item Choose a probability distribution $\probdist^{(t)}_\PP$ on
  $\PP$.
\item Obtain points $\vec{g}_1^{(t)},\dots,\vec{g}_{\lambda}^{(t)}$ by
  taking $\lambda$ samples from the distribution
  $\probdist^{(t)}_\PP$.\label{sec:step}
  Evaluate $\Phi$ (perhaps with random noise) at these points.
\item According to a fixed (algorithm dependent) rule construct a
  probability distribution $\probdist^{(t+1)}_\PP$ on $\PP$.\label{thr:step}
\item Check for some appropriate stopping condition; if the algorithm has
  not terminated, substitute $t:=t+1$ and return to Step \ref{sec:step}.
\end{enum}
\noindent The core ingredient of this search scheme is the search distribution
$\probdist^{(t)}_\PP$, also called exploration distribution, on the
search space.  Random search algorithms can differ fundamentally in
the way they represent and alter the search distribution.  Typically,
the distribution is represented by a semi-parametric model. The choice
of the model determines the exploration distributions the search
algorithm can represent, which is in general only a small subset of
all possible probability distributions on $\PP$.  For example, let
$\PP = \RRR$. Then the class of representable distributions may be
given by the class of normal densities, $p_\PP(x;m,\sigma) =
\frac{1}{\sqrt{2\pi}\sigma} \exp\left\{ -\frac{(x-m)^2}{2\sigma^2}
\right\}$, where $m$ is the expectation and $\sigma^2$ the variance.
This equation defines a \emph{parameterization} of the search
distribution, it maps the parameters $(m,\sigma)\in\RRR^2$ into the
set of probability distributions on $\PP$ (see Fig.~\ref{huete:fig}).
A global random search algorithm alters its exploration distribution
(see Step \ref{thr:step}) by changing such parameters.

We believe that two of the most fundamental questions in evolution
theory can be stated as:
\begin{enum}
\item How is the exploration distribution parameterized?
\item How is the exploration distribution altered?
\end{enum}
\noindent 
In the framework of evolutionary systems, we identify each element of
$\PP$ with certain phenotypic traits of individuals and call $\PP$ the
phenotype space. Each phenotype $p$, i.e., element of $\PP$, is
encoded by a genotype $\vec{g}$, which is an element of a genotype
space $\GG$. The mapping $\gp:\GG\to\PP$ is called genotype-phenotype
mapping.  If $\gp$ is not injective we speak of a \emph{neutral}
encoding.  A number of genotypes
$\tilde{\vec{g}}^{(t)}_1,\dots,\tilde{\vec{g}}^{(t)}_\mu$, the
parents, are stored in a population. The superscript indicates the
iteration of the algorithm, i.e., the generation.  In each generation,
$\lambda$ offspring $\vec{g}^{(t)}_1,\dots,\vec{g}^{(t)}_\lambda$ are
generated by applying stochastic and\,/\,or deterministic variation
operators.  Let the probability that parents
$\tilde{\vec{g}}_1,\dots,\tilde{\vec{g}}_\mu\in\GG$ generate an
offspring $\vec{g}\in\GG$ be described by the probability distribution
$P_\GG\left(\vec{g} \;|\; \tilde{\vec{g}}_1, \dots,
\tilde{\vec{g}}_\mu; \vec{\theta}\right)$.  This distribution is
additionally parameterized by some \emph{external strategy parameters}
$\vec{\theta}\in\Theta$.  Examples of such externally controlled
parameters include the probabilities that certain variation operators
are applied and parameters that determine the mutation strength.  We
call the probability
\begin{align}
P_\GG^{(t)}(\vec{g})
  := P_{\GG}\left(\vec{g} \;|\; \tilde{\vec{g}}_1^{(t)}, \dots,
   \tilde{\vec{g}}^{(t)}_\mu ; \vec{\theta}^{(t)}\right)
\end{align}
that in generation $t$ an offspring $\vec{g}$ is created
the \emph{exploration distribution} on $\GG$ at generation $t$.

The genotype-phenotype mapping $\gp$ \emph{lifts} $P_{\GG}^{(t)}$ from
the genotype space onto the phenotype space:
\begin{align}\label{neuset}
\forall p\in \PP:~ \EE^{(t)}(p)
 = \sum_{\vec g' \in \gp^{-1}(p)} P_\GG^{(t)}(\vec g') \;,
\end{align}
where $\gp^{-1}(p):=\{\vec g' \in \GG \;|\; \gp(\vec g')=p\}$ is
called the \emph{neutral set} of $p\in \PP$ \cite{schuster:96}. Thus,
the genotype space $\GG$ together with the genotype-phenotype mapping
$\gp$, the variation operators, and external parameters $\vec\theta$
can be regarded as a parameterization of the exploration distribution
on the search space $\PP$.  We refer to \cite{toussaint:01} for more
details on this way of formalizing evolutionary exploration.

Also algorithms recently developed in the field of evolutionary
computing can be captured by this point of view: M\"uhlenbein et
al.~\cite{muehlenbein:99} and Pelikan et al.~\cite{pelikan:00}
parameterize the exploration density by means of a Bayesian dependency
network in order to introduce correlations in the exploration
distribution. The CMA evolution strategy proposed by Hansen and
Ostermeier \cite{hansen:01} adapts a covariance matrix that describes
the dependencies between real-valued variables in the exploration
density. In the remainder of this paper, we focus on a different
approach that might be considered biologically more plausible, namely
to utilize an appropriate genotype-phenotype mapping to parameterize
the exploration density $\EE$. It does not explicitly encode
correlations but models such interactions indirectly via the
genotype-phenotype mapping. What we will ask for in the next section
is the way it allows for self-adaptation of the exploration strategy.




\section{Self-adaptation}\label{selfada}

\subsection{Introduction}\label{selfintro}

The ability of an evolutionary algorithm to adapt its search strategy
during the optimization process is a key concept in evolutionary
computation, see the overviews \cite{angeline:95,smith:97,eiben:99}.
Online adaptation of strategy parameters is important, because the
best setting of an EA is usually not known \emph{a priori} for a given
task and a constant search strategy is usually not optimal during the
evolutionary process. One way to adapt the search strategy online is
self-adaptation, see \cite{baeck:98} for an overview.  This method can
be described as follows \cite{eiben:99}: ``The idea of the evolution
of evolution can be used to implement the self-adaptation of
parameters. Here the parameters to be adapted are encoded into the
chromosomes and undergo mutation and recombination. The better values
of these encoded parameters lead to better individuals, which in turn
are more likely to survive and produce offspring and hence propagate
these better parameter values.''  In other words, each individual
not only represents a candidate solution for the problem at hand, but
also certain strategy parameters that are subject to the same
selection process---they hitchhike with the object parameters.
 
Self-adaptation is used in all main paradigms of evolutionary
computation.  The most prominent examples stem from evolution
strategies, where it is used to adapt the covariance matrix of the
mutation operator, see Sec.~\ref{es:sec}.  Self-adaptation is
employed in evolutionary programming for the same purpose, but also in
the original framework of finite-state machines \cite{fogel:95b}.  In
genetic algorithms, the concept of self-adaptation has been used to
adapt mutation probabilities \cite{baeck:92b} and crossover operators
\cite{schaffer:87}.  Self-adaptive crossover operators have also been
investigated in genetic programming \cite{angeline:96b}. In the
following, we will propose an alternative definition of
self-adaptation, which includes these approaches as special cases.

\subsection{Neutrality and self-adaptation}\label{neu}

Following the formalism we developed in the previous section, a
variation of the exploration density $\EE^{(t)}$, corresponding to
Step \ref{thr:step} in the global random search algorithm, occurs if
\begin{enum}
\item the parent population $(\tilde{\vec g}_1^{(t)}, \dots,
  \tilde{\vec g}_\mu^{(t)})$ varies, or
\item the external strategy parameters $\vec\theta^{(t)}$ vary, or
\item the genotype-phenotype mapping $\gp$ varies.
\end{enum}
\noindent 
From a biological point of view, one might associate environmental
conditions (like the temperature, etc.) with external parameters that
vary the mutation probabilities or the genotype-phenotype mapping. In
most cases one would not consider them as subject to
evolution.\footnote{Counter-examples that go far beyond the scope of
  our formalism are, for instance, the embryonic environment (uterus)
  and the inherited ovum, or individuals whose behavior have an
  influence on mutations (e.g., sexual behavior influencing crossover)
  or on the embryonic development (e.g., a mother taking drugs). All
  of these influences might be considered as subject to evolution.  In
  particular the interpretation of an ovum is a critical issue.
  Should one regard it as part of the genotype or as an inherited
  ``parameter'' of the genotype-phenotype mapping?} Sometimes
mechanisms that adapt the exploration distribution by genotype
variations are regarded as examples of adaptive genotype-phenotype
mappings. To our minds, this is a misleading point of view.  For
example, \mbox{t-RNA} determines a part of the genotype-phenotype
mapping and it is itself encoded in the genome.  However, the
genotype-phenotype mapping should be understood as a whole---mapping
all the genotype (including the parts that code for the t-RNA) to the
phenotype, such that it becomes inconsistent to speak of genotypic
information parameterizing the genotype-phenotype mapping.

The same arguments also apply in the context of evolutionary
computation and thus we consider only option 1 as a possibility to
vary the exploration distribution in a way that is itself subject to
evolution in the sense of section \ref{selfintro}. However, if the
genotype-phenotype mapping is injective, every variation of genotypes
alters phenotypes and bears the risk of a fitness loss. Hence, we
conclude:

\smallskip \emph{In the absence of external control, only neutral
  genetic variations can allow a self-adaptation of the exploration
  distribution without changing the phenotype, i.e., without the risk of
  loosing fitness.} \smallskip

\begin{figure}[t]\center
  \input{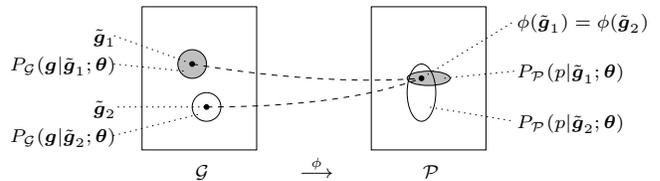}
\caption{
  Two different points $\tilde{\vec g}_1,\tilde{\vec g}_2$ in $\GG$
  are mapped onto the same point in $\PP$. The elliptic ranges around
  the points illustrate the exploration distributions by suggesting
  the range of probable mutants.  Thus, the two points $\tilde{\vec
    g}_1,\tilde{\vec g}_2$ belong to one neutral set but represent two
  different exploration strategies on the search space $\PP$.}
\label{strategy}
\end{figure}

\noindent A \emph{neutral genetic variation} means that parent and offspring
have the same phenotype. For instance, consider two genotypes
$\tilde{\vec g}_1$, $\tilde{\vec g}_2$ in a neutral set. Neglect
crossover and assume that the probability for an offspring $\vec g$ of
a single parent $\tilde{\vec g}_i$ is given by $P_\GG(\vec g \;|\;
\tilde{\vec g}_i;\vec\theta)$. The two genotypes may induce two
arbitrarily different exploration distributions ``around'' the same
phenotype $p=\gp(\tilde{\vec g}_1)=\gp(\tilde{\vec g}_2)$, see
Fig.~\ref{strategy}. Transitions between these genotypes allow for
switching the exploration strategy. In general, the variety of
exploration densities that can be explored in a neutral set
$\gp^{-1}(p)$ is given by
\begin{align}
\big\{ P_\PP(p \;|\; \tilde{\vec g}_i;\vec\theta) \;\big|\; \tilde{\vec g}_i \in
 \gp^{-1}(p) \big\} \;.
\end{align}

\subsection{Discussion}

It is widely accepted that changing the genotypes without
significantly changing the phenotypes in the population is a key
search strategy in natural evolution
\cite{huynen:96b,kimura:68,schuster:02}. We emphasize that under the
stated assumptions, neutrality is even a necessity for self-adaptation
of the search strategy. Thus, we propose to \emph{define
  {self-adaptation} as the use of neutrality in order to vary the
  exploration strategy}.

Existing approaches to self-adaptation developed in the realm of
evolutionary computation can be embedded in this point of view; only
the \emph{style} in which the notion of self-adaptation was originally
introduced is different. Usually one refers to ``strategy parameters''
that typically control the mutation operators (i.e., the mapping
$\tilde{\vec g}_i \mapsto P_\GG(\vec g \;|\; \tilde{\vec
  g}_i;\theta)$), in contrast to ``object parameters'' describing
fitness relevant traits.  In the case of \emph{self}-adaptation these
strategy parameters are considered as parts of the genotype. In this
view, strategy parameters are neutral, i.e., altering them does not
change the phenotype.  In turn, we regard neutral genotypic variables
as potential strategy parameters. There still is a slight difference:
In case of standard self-adaptation, altering a strategy parameter is
phenotypically neutral and has no other implication than a change of
the exploration strategy. In contrast, two or more genetic variables
may be phenotypically non-neutral but a specific simultaneous mutation
of them might be neutral and induce a change of the exploration
density.  Hence, a neutral set is a general concept for
self-adaptation, which is not bound to the idea of single loci being
responsible for the search strategy. A good example is that also
topological properties of the search space (i.e., neighborhood
relations/the set of most probable mutations of a phenotype) may vary
along such a neutral set if the genotype-phenotype mapping is chosen
similar to grammars \cite{toussaint:01}---which seems hard to realize
by explicit strategy parameters.

We believe that allowing for a self-adaptive exploration strategy is
the main benefit of neutral encodings, as stated in
\cite{shackleton:00}: ``For redundancy to be of use it must allow for
mutations that do not change the current phenotype, thus maintaining
fitness, and which allow for moves to areas where new phenotypes are
accessible''.  Changing the exploration distribution corresponds to
the ability of reaching new phenotypes.\footnote{However, we think
  that there might be at least one additional case where neutrality
  can be useful, namely when it introduces---\emph{by chance}---a bias
  in the search space, such that desired phenotypes are (in global
  average) represented more often than other ones \cite{igel:01d}.}

In this view, neutrality is not necessarily \emph{redundant}:
Different points in a neutral set can encode different exploration
strategies and thus different information; a genotype encodes not only
the information on the phenotype but also information on further
explorations. One might state it like this: Although a non-injective
genotype-phenotype mapping can be called redundant, the corresponding
mapping from genotype to exploration distribution may in general be
non-redundant.

\section{Examples}\label{examples}

\subsection{Codon Bias in natural evolution}

An intriguing study of the interrelationship between neutrality and
self-adaptation in nature is the one by Stephens and Waelbroeck
\cite{stephens:99}. They empirically analyze the codon bias and its
effect in RNA sequences of the HI virus. In biology, several
nucleotide triplets eventually encode the same amino acid. For
example, there are 9 triplets that are mapped to the amino acid
Arginine. If Arginine is encoded by the triplet CGA, then the chance
that a single point mutation within the triplet is neutral (does not
chance the encoded amino acid) is $4/9$. In contrast, if it is encoded
by AGA, then this \emph{neutral degree} is $2/9$. Now, codon bias
means that, although there exist several codons that code for the same
amino acid (which form a neutral set), HIV sequences exhibit a
preference on which codon is used to code for a specific amino acid.
More precisely, at some places of the sequence codons are preferred
that are ``in the center of this neutral set'' (with high neutral
degree) and at other places codons are biased to be ``on the edge of
this neutral set'' (with low neutral degree).  It is clear that these
two cases induce different exploration densities; the prior case means
low mutability whereas the latter means high mutability. Stephens and
Waelbroeck go further by giving an explanation for these two
(marginal) exploration strategies: Loci with low mutability cause
``more resistance to the potentially destructive effect of mutation'',
whereas loci with high mutability might induce a ``change in a
neutralization epitope which has come to be recognized by the
[host's] immune system'' \cite{stephens:99}.


\subsection{Self-adaptation in evolution strategies}\label{es:sec}

In this section, we describe a rather simple example of
self-adaptation in evolution strategies.\footnote{More sophisticated
  and efficient algorithms exist for dealing with real-world
  optimization tasks, e.g., the derandomized adaptation as proposed
  in \cite{hansen:01,ostermeier:95}.} The candidate solutions for the
problem at hand are represented by real-valued vectors $\vec{x}\in
{\mathcal P}={\mathcal G}\subseteq{\mathbbm R}^n$. These real-valued
object parameters are mutated by adding a realization of a normally
distributed random vector with zero mean
\cite{rechenberg:94,schwefel:95}. The symmetric $n\times n$ covariance
matrix of this random vector is subject to self-adaptation.  To
describe any possible covariance matrix, $n(n+1)/2$ strategy
parameters are needed. However, often a reduced number of parameters
is used. In the following, let the covariance matrix be given by
$(\vec{\sigma} I)^2$, where $I=\operatorname{diag}(1,\dots,1)$ is the
identity matrix and $\vec{\sigma}\in{\mathbbm R}^n$ corresponds to $n$
strategy parameters that describe the standard deviations along the
coordinate axes.

Let $\mu$ denote the size of the parent population.  Each generation,
$\lambda$ offspring $\vec g_i^{(t)}$, $i=1,\dots,\lambda$ are
generated.  $(\mu,\lambda)$-selection is used, i.e., the $\mu$ best
individuals of the offspring form the new parent population.
An individual $\vec g_i^{(t)}
\in {\mathbbm R}^{2n}$ can be divided into two parts, $\vec g_i^{(t)}
= (\vec x_i^{(t)} , \vec \sigma_i^{(t)})$, the object variables $\vec
x_i^{(t)} \in {\mathbbm R}^n$ and the strategy parameters $\vec
\sigma_i^{(t)} \in {\mathbbm R}^n$.

For each offspring $\vec{g}_i^{(t)}$, two indices
$a,b\in\{1,\dots,\mu\}$ are selected randomly, where $a$ determines
the parent and $b$ its mating partner.  For each component
$k=1,\dots,n$ of the offspring, the new standard deviation $\vec
\sigma_i^{(t)}=(\sigma_{i,1}^{(t)},\dots,\sigma_{i,n}^{(t)})$ is
computed as
\begin{equation}\label{sigma:eq}
\sigma_{i,k}^{(t)}
 = \frac{1}{2}\left[\tilde{\sigma}_{a,k}^{(t)}+\tilde{\sigma}_{b,k}^{(t)} \right]
 \cdot \exp\left(\tau' \cdot \xi_i^{(t)} + \tau \cdot \xi_{i,k}^{(t)}\right)
\enspace.
\end{equation}
Here, $\xi_{i,k}^{(t)} \sim \zz{N}({0}, 1)$ is a realization of a
normally distributed random variable with zero mean and variance one
that is sampled anew for each component $i$ for each individual,
whereas $\xi_i^{(t)} \sim \zz{N}({0}, 1)$ is sampled once per
individual and is the same for each component.  The log-normal
distribution ensures that the standard deviations stay positive. For
the mutation strengths $\tau\propto 1/\sqrt{2\sqrt{n}}$ and
$\tau'\propto 1/\sqrt{2{n}}$ are recommended
\cite{schwefel:95,baeck:98}.  It has been shown that recombination of
the strategy parameters is beneficial (e.g., \cite{kursave:99}).
Equation (\ref{sigma:eq}) realizes an intermediate recombination of
the strategy parameters.

\begin{figure}
\small
\input cec.frag
\hfill\epsfig{file=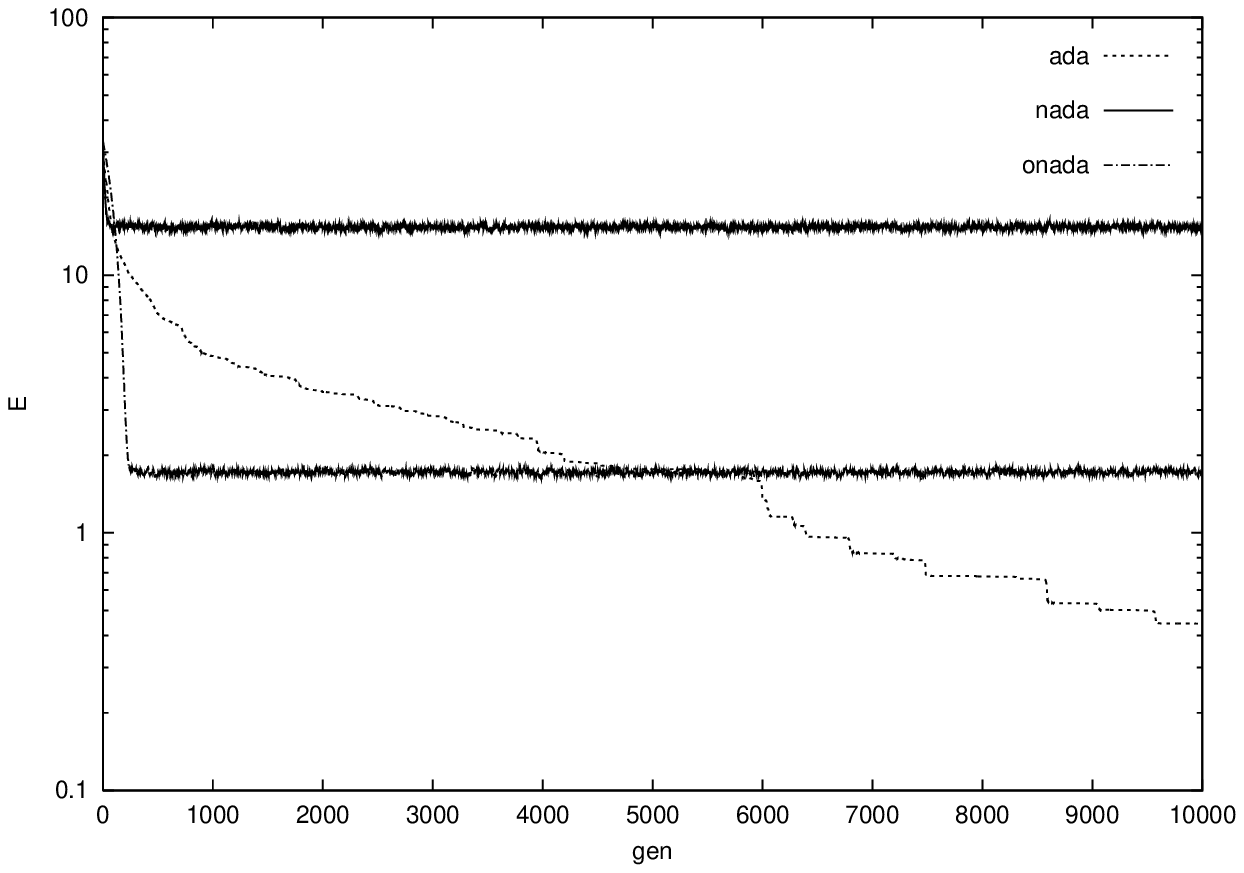, width=.9175\columnwidth}\hspace{1ex}

\hfill\epsfig{file=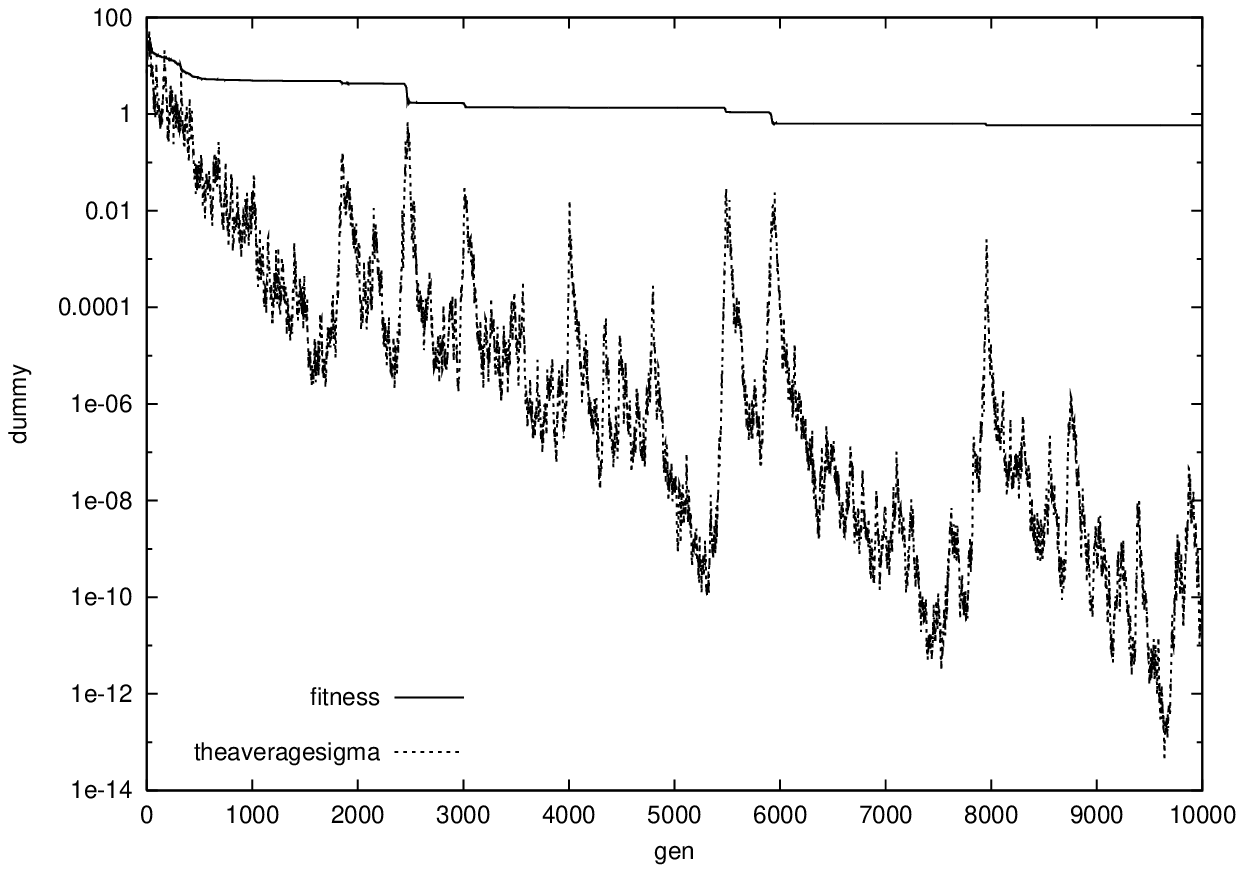, width=.95\columnwidth}\hspace{1ex}

\caption{\label{average:fig}As an example, we compare
  evolution strategies as described in Sec.~\ref{es:sec} with and
  without self-adaptation on a simple test problem: a 100-dimensional
  sphere model, $\Phi(\vec{x})=||\vec{x}||$.  First, the EA is run
  without self-adaptation, each $\sigma_i$ is set to one.  After that,
  we employ self-adaptation as described, where the $\sigma_i$ are
  also initialized to one.  Based on these results, we determine the
  average standard deviation $\sigma^\dagger$ (averaged over all
  parents, all generations, and all trials).  Then we repeat the
  experiments without
  self-adaptation but fix each $\sigma_i=\sigma^\dagger$.
  The fitness trajectories of the best individual averaged over 50
  trials are shown in the upper plot ($\sigma^\dagger=0.0125$). 
  In case of the self-adaptive EA the fitness curves show steps.  The
  lower plot shows a typical single trial and the corresponding
  average of the standard deviations
  $||\left<\vec{\sigma}\right>_\mu^{(t)}||= \Big(\sum_{j=1}^n\left(
  \frac{1}{\mu}\,\sum_{i=1}^\mu \sigma_{i,j} \right)^2\Big)^{1/2}$.
  It becomes obvious that each step in the fitness trajectory
  corresponds to a change in the search strategy. In general, the
  mutation strength decreases, but sometimes a larger step size takes
  over the population and leads to a high
  fitness gain.
  In the continuum, the
  probability that an offspring has exactly the same phenotype, i.e.,
  represents the same object parameters, is not measurable. Further,
  even small changes in the genotype are relevant for selection due to
  the use of a rank-based selection scheme.  Hence, \emph{drifting}
  along a \emph{neutral network}, i.e., traversing the search space by a
  sequence of (\emph{neutral}) mutations that do not alter the
  phenotype, does not occur.
 }
\end{figure}

Thereafter the objective parameters 
are altered using the new strategy parameters:
\begin{equation}
x_{i,k}^{(t)}
 = \tilde{x}_{a,k}^{(t)} + \sigma_{i,k}^{(t)}\, z_{i,k}^{(t)},
\end{equation}
where $z_{i,k}^{(t)}\sim\zz{N}({0}, 1)$.  For simplicity, we do not
employ recombination of the objective parameters.  An example of this
strategy compared to algorithms without self-adaptation is shown in
Fig.~\ref{average:fig}.  The adaptive strategy performs considerably
better than the other methods, as known from many theoretical results
and empirical studies. However, we would like to underline that the
self-adaptive EA uses a highly redundant encoding instead of a
one-to-one genotype-phenotype mapping---the genotype space has twice
the dimensionality of the phenotype space. The neutrality does not
alter the distribution of the fitness values, it just allows the
search strategy to adapt.

\section{Conclusion}

Neutrality is a necessity for self-adaptation. Actually, the design of
neutral encodings to improve the efficiency of evolutionary algorithms
is a well-established approach: strategy parameters are an example of
neutrality. Hence, there already exists clear evidence for the benefit
of neutrality.

The notion of neutrality provides a unifying formalism for embedding
approaches to self-adaptation in evolutionary computation. But it also
inspires new approaches that are not bound to the idea of only single
genes being responsible for the exploration strategy.  Generally, any
local properties of the phenotypic search space---metric or
topological---may vary along a neutral set. An example are
grammar-based genotype-phenotype mappings, for which different points
in a neutral set represent completely different topologies (and thus
exploration strategies) on the phenotype space.

On the other hand, the benefits of neutrality in {natural}
evolution can be better understood when the simple---but concrete and
well-established---paradigms of self-adaptation and neutrality in
evolutionary computation are taken into account. As we exemplified
with the second example, these computational approaches may serve as a
model to investigate the relation between evolutionary progress,
neutrality, and self-adaptation.

Finally, let us recall that neutrality is not necessarily equivalent
to redundancy: in general, a genotype encodes not only the information
on the phenotype but also information on further explorations.

\subsection*{Acknowledgments}
The authors acknowledge support by the
DFG under grant Solesys SE 251/41-1.

{ \small 
\bibliographystyle{abbrv} 
\bibliography{cec02,phd,mt} 
}
\end{document}